\documentclass[article]{elsarticle}

\usepackage{lineno,hyperref}
\usepackage{multirow}
\usepackage{array}
\usepackage{color}
\usepackage{xcolor}%
\usepackage{soul}
\begin{document}
% \linenumbers

\begin{frontmatter}

\title{
    Evaluation of cosmogenic Ge-68 background in a high purity germanium detector 
    via a time series fitting method 
}

\author[THU]{W.H. Dai}
\author[THU]{J.K. Chen}
\author[THU]{H. Ma\corref{mycorrespondingauthor}}
\cortext[mycorrespondingauthor]{Corresponding author: mahao@tsinghua.edu.cn}
\author[THU]{Z. Zeng}
\author[THU]{M.K. Jin}
\author[THU]{Q.L Zhang}
\author[THU,BNU]{J.P. Cheng}

\address[THU]{Key Laboratory of Particle and Radiation Imaging 
(Ministry of Education) and Department of Engineering Physics, 
Tsinghua University, Beijing 100084}
\address[BNU]{College of Nuclear Science and Technology, 
Beijing Normal University, Beijing 100875, China}

\begin{abstract}
Ge-68 is a cosmogenic isotope in germanium with a half-life of 270.9 days.
Ge-68 and its decay daughter Ga-68 contribute considerable
background with energy up to 3 MeV to low background $\gamma$ spectrometers
using high purity germanium (HPGe) detectors.
In this paper, we evaluated the background of Ge-68 and Ga-68
in a $p$-type coaxial HPGe detector operated at China Jinping underground laboratory (CJPL)
via a time series fitting method. 
Under the assumption that Ge-68 and Ga-68 are in radioactive equilibrium and 
airborne radon daughters are uniformly distributed in the measurement chamber of the spectrometer,
we fit the time series of count rate in 1-3 MeV 
to calculate the Ge-68 activity, radon daughter concentrations,
and the time-invariant background component.
A total of 90-day measurement data were used in the analysis, 
a hypothesis test confirmed a significant Ge-68 signal at 99.64\% confidence level.
The initial activity of Ge-68 is fitted to be 477.0$\pm$112.4 $\mu$Bq/kg, 
corresponding to an integral count rate of 55.9 count/day in the 1-3 MeV range.
During the measurement, Ge-68 activity decreased by about 30\%, 
contributing about 62\% of the total background in the 1-3 MeV range.
Our method also provides an estimation of the variation of airborne radon daughter concentrations
in the measurement chamber, 
which could be used to monitor the performance of radon reduction measures.
\end{abstract}

\begin{keyword}
    high purity germanium detector, cosmogenic Ge-68, time series analysis
% \MSC[2010] 00-01\sep  99-00
\end{keyword}

\end{frontmatter}

% \linenumbers

\section{Introduction}{\label{sec.1}}
High purity germanium (HPGe) detectors have been widely used in 
radiation monitoring, nuclear physics, particle physics, and astrophysics 
due to its high energy resolution, high stopping power, and low intrinsic background 
\cite{bib:1,bib:2,bib:3}.
The use of an HPGe detector as $\gamma$ spectrometer requires strict control and accurate measurement of background 
for detecting trace radioactivity within the sample.
For $\gamma$ spectrometers operated at underground laboratories,
the rock overburden reduces the cosmic-ray muon flux by several orders of magnitudes, 
for instance, 1000 m rock can reduce cosmic-ray muon flux by over $10^5$ times
\cite{bib:4}.
External $\gamma$ can be shielded by high-Z materials like copper and lead,
neutron can be shielded by borated polyethylene.
However, the cosmogenic radioactive isotopes cumulated during the fabrication and transportation of 
the detector above ground can continuously contribute to the background after the detector has been moved underground.
As the production of cosmogenic isotopes becomes negligible underground \cite{bib:5},
their activities decrease after the detector arrives underground,
leading to a change of background in the measured spectrum.
Therefore, measuring the cosmogenic isotope activities and evaluating their background
in an underground HPGe spectrometer is important for the background modeling and understanding.

\begin{figure}[!htb]
    \centering
    {\includegraphics[width=0.47\linewidth]{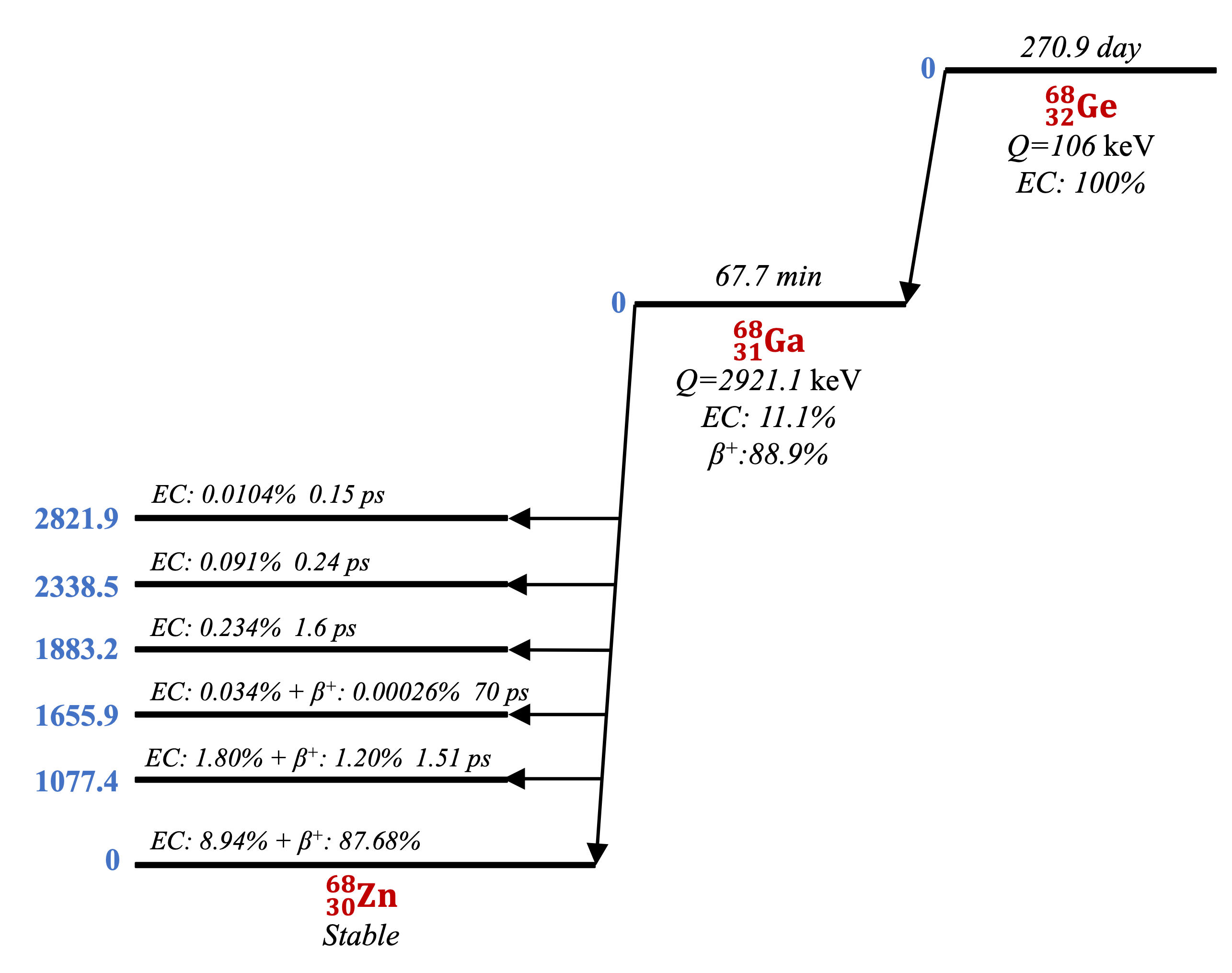}}
    {\includegraphics[width=0.47\linewidth]{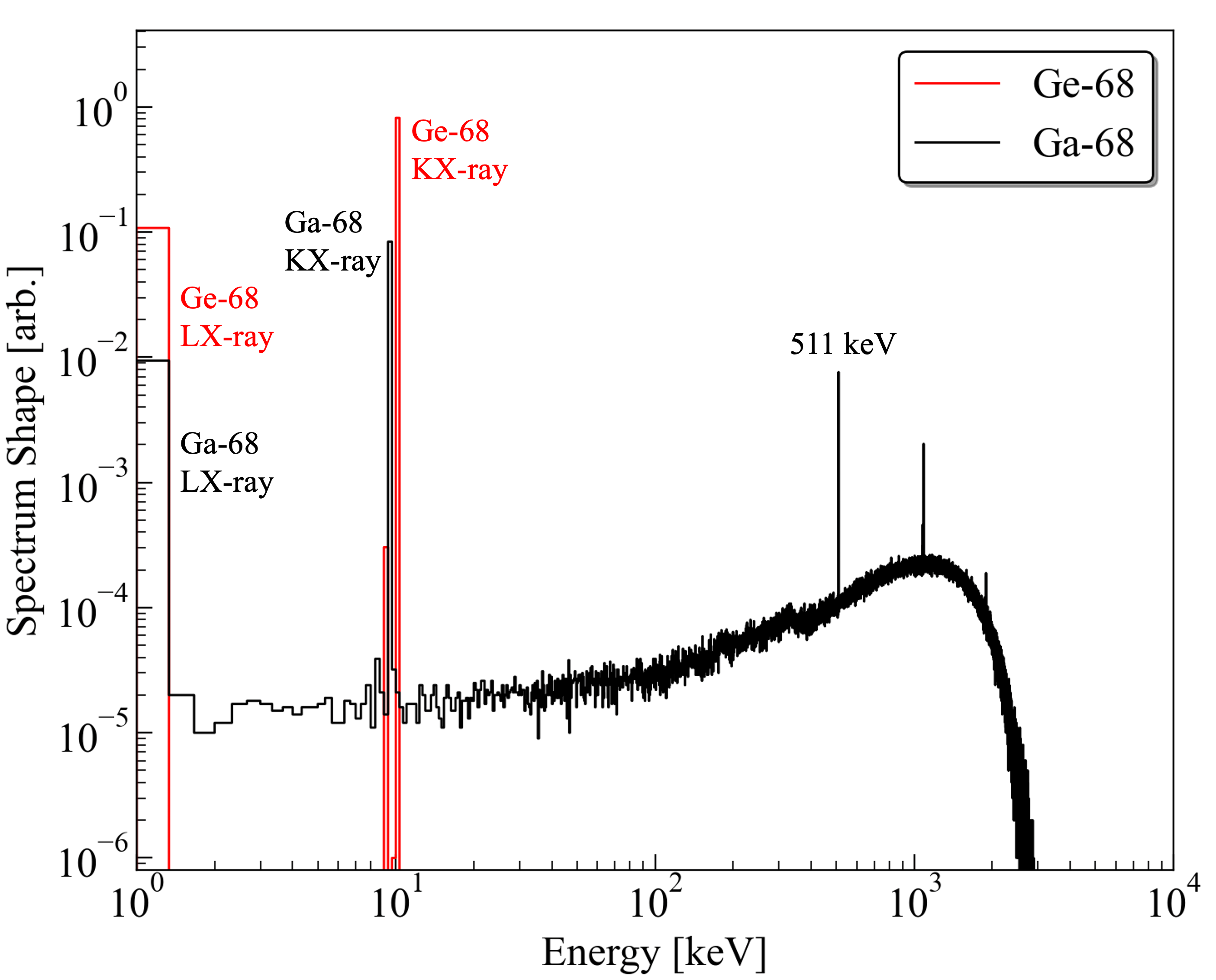}}
    \caption{\label{fig:1} Left panel: decay scheme of Ge-68 and Ga-68, data from\cite{bib:6}. 
    Right panel: typical Ge-68 and Ga-68 spectra in an HPGe detector, the spectra are simulated via Geant4\cite{bib:7,bib:8,bib:9} 
    using a 1-kg cylinder Ge model with 0.5 mm dead layer at the surface (energy resolution not considered).}
\end{figure}

Ge-68 is a cosmogenic isotope produced by nuclear reactions between the germanium nucleus and
high energy cosmic rays (neutron, proton, and $\gamma$).
At sea level, the production rate is around 80-120 $\rm kg^{-1}day^{-1}$ for different latitudes and longitudes
\cite{bib:10},
corresponding to a saturated activity of 0.93-1.39 mBq/kg.
Ge-68 decays to Ga-68 via electron capture (EC) with a half-life of 270.9 days (decay scheme in Fig.\ref{fig:1})
\cite{bib:6}.
Its decay emits X-rays and auger electrons with total energy equaling to the binding energy of the captured electron ($<$11 keV),
therefore only contributing background at low energy region (as in Fig.\ref{fig:1}(right)).
Ga-68 decays to Zn-68 with a $Q$-value of 2921.1 keV and a half-life of 67.7 min
\cite{bib:6},
its decay scheme is shown in Fig.\ref{fig:1}(left).
In most cases, Ga-68 decays to the ground state of Zn-68 via $\beta^+$ mode and emits a positron
with a maximum kinetic energy up to 1899.1 keV.
The spectrum shape of Ga-68 in a cylindrical HPGe detector is shown in Fig.\ref{fig:1}(right),
it manifests an arc-shaped line with a small peak from 511 keV annihilation photons
and summed peaks around 1 MeV from  $\gamma$ and X-rays in the EC decay mode.
Due to the relatively short half-life of Ga-68, it will be in radioactive equilibrium with Ge-68.
Ga-68 with an activity of 1 mBq/kg could contribute a background of approximately 85 count/day/kg$\rm _{Ge}$ at 60-2700 keV range,
which is comparable to the typical background level of an underground HPGe $\gamma$-spectrometer
\cite{bib:11}.
As the energy of Ge-68 background spans from keV to MeV,
it is also an important background both in dark matter direct search experiments
(SuperCDMS\cite{bib:ad-2} and CDEX\cite{bib:ad-3})
and neutrinoless double beta decay experiments
(GERDA\cite{bib:ad-4}, Majorana\cite{bib:ad-5}, LEGEND\cite{bib:ad-6}).
The measurement of the Ge-68 activity is important for the 
background understanding of those low background experiments\cite{bib:10,bib:ad-II-3,bib:ad-II-5}.

There are general two ways of measuring the Ge-68 activity, 
one is analyzing the 10.38 keV KX-ray peak in the spectrum
\cite{bib:10,bib:12,bib:13}.
However, this method is not feasible for detectors with a dynamic range that cannot cover the low energy  range.
The other way is fitting the count rate at different times.
Once the detector arrives underground, 
the decay of Ge-68 and Ga-68 will decrease the count rate in the spectrum.
Fitting the count rate with Ge-68 half-life could provide an estimation of the activity,
but requires decoupling other time variant components, 
e.g., the count rate change caused by concentration variation of the airborne radon daughter.
We develop a method fitting the count rate with the Ge-68 decay component and
the variation of radon daughter simultaneously 
while constraining the radon daughter concentration by its characteristic peaks.
This method is then applied to measuring the activity of Ge-68 in an HPGe detector 
operated in China Jinping underground laboratory (CJPL)\cite{bib:14}.

This paper is organized as follows: 
Sec.\ref{sec.1} gives the background of this work.
Sec.\ref{sec.2} introduces the detector, fitting method and statistic test method. 
Sec.\ref{sec.3} provids the fitting results and interpretation of the results.
Sec.\ref{sec.4} summaries the this work and outlooks the possiable applications of our method.

\section{Method}
\subsection{HPGe detector at CJPL}
\begin{figure}[!htb]
    \centering
    {\includegraphics[width=1.0\linewidth]{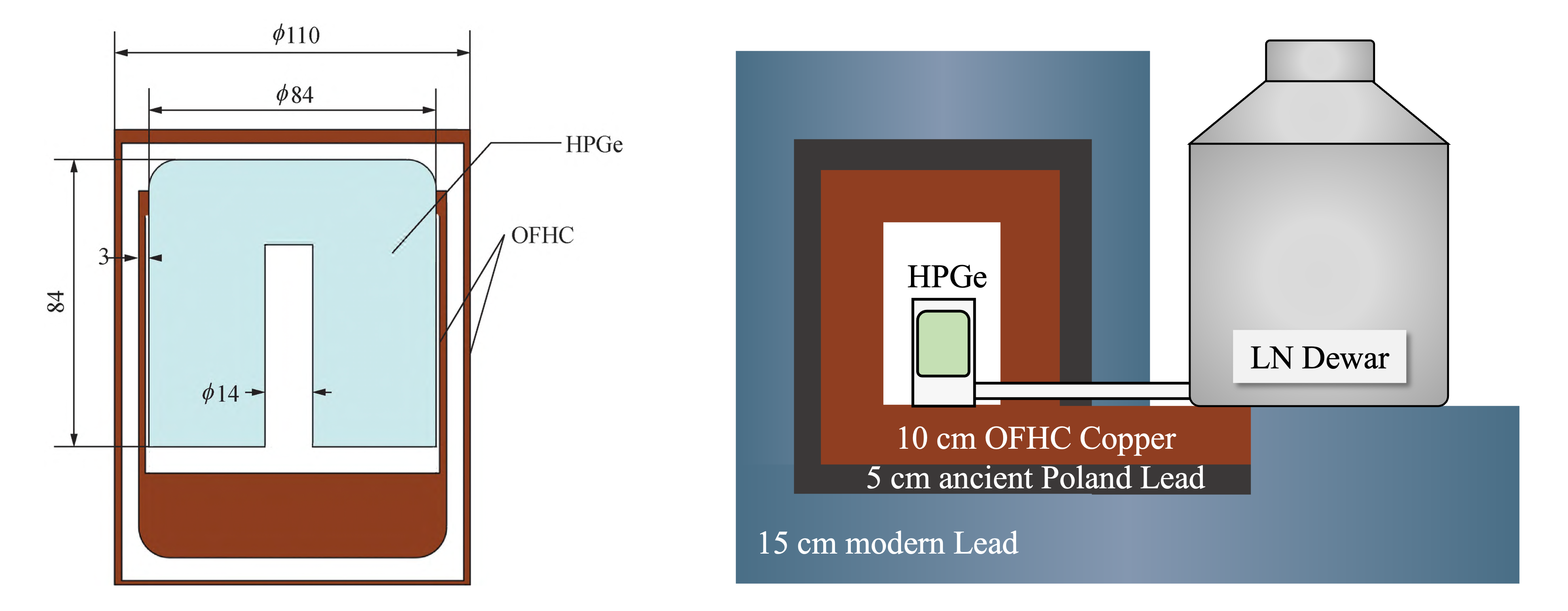}}
    \caption{\label{fig:2} Left panel: structure of the HPGe detector.
    Right panel: structure of the detector's shielding, the shielding consists of
    innermost 10 cm OFHC copper, 5 cm ancient Poland lead, and outmost 15 cm modern lead.}
\end{figure}

{\label{sec.2}}
The HPGe detector studied is a low background $p$-type coaxial HPGe detector purchased from CANBERRA
and is used as the detector of a low background $\gamma$-spectrometer at CJPL.
It has a Ge crystal of 2.48 kg with 0.5 mm dead layer and an energy threshold of 100 keV,
the crystal and its surrounding structure are shown in the left panel of Fig.\ref{fig:2} 
\cite{bib:15}.
After the detector had been manufactured in France, it was shipped to CJPL via truck and train,
it arrived at CJPL on 2020/07/25 with an approximate 1 month exposure time above ground 
\cite{bib:15}.

After arriving at CJPL, the detector is shielded by copper and lead to reduce the 
background from environmental $\gamma$-rays.
The shielding is made of 10 cm underground storage copper 
and 20 cm lead (5 cm ancient Poland lead and 15 cm modern lead), 
the structure of the Cu/Pb shielding is illustrated in Fig.\ref{fig:2}.
Nitrogen gas has been constantly injected into the detector chamber inside the copper shielding
to reduce the radon concentration.

The energy calibration was performed using Co-60, Co-57, and Eu-152\cite{bib:15}.
After the calibration, the sources were removed and the detector was operated to measure the background.
The spectrum was saved regularly, each saved spectrum corresponds to 6 hours of live measurement time.
The data taken started in 2020/08/15 and ended in 2020/11/21, 
a total of 90 days live time background measurement data is cumulated and
used to calculate the activities of cosmogenic isotopes.
The measurement of Mn-54, Co-57, and Co-58 activities via fitting their characteristic peaks could be found in\cite{bib:15}.
This work focuses on the measurement of Ge-68 activity.

\subsection{Time series fitting method}{\label{sec.2.1}}
The calculation of Ge-68 activity via fitting the variation of count rate requires
decoupling various time variant and invariant background.
The time-variant backgrounds considered are cosmogenic isotopes with relatively short half-lives
and radon daughters in the void volume between the detector and copper shield.
The time-invariant background are those from radioisotopes with long half-lives,
for instance, $\gamma$-rays from U, Th decay chain, and K-40.
And for background from Co-60,
as the half-life of Co-60 (1925.3 days) is much longer than the measurement time (90 days), 
its activity decreases only 3\% during the measurement.
Therefore, it is treated as a time-invariant background in this work.
The count rate in a specified energy range can be modeled as Eq.\ref{eq:1}

\begin{eqnarray}
    \label{eq:1}
    R_i(t,t+\Delta t)=&& 
    \sum_{k}^{N_{Iso}}\frac{A_{0,k}}{\Delta t}\cdot \frac{T_{1/2,k}}{\ln 2}\cdot e^{-\frac{\ln 2}{T_{1/2,k}}t}
    [1-e^{-\frac{\ln 2}{T_{1/2,k}}\Delta t}]  \cdot \varepsilon_{k,i} \\ \nonumber
    &&+\sum_{p}^{N_{Rn}}C_{p}(t)\cdot \varepsilon_{p,i} + B_i,
\end{eqnarray}

\noindent 
the specified energy range is indicated by $i$,
$R_i(t,t+\Delta t)$ is the count rate in $i$ energy range at $t\sim t+\Delta t$ interval.
The first item in Eq.\ref{eq:1} is the sum of the contribution from each cosmogenic isotope $k$,
$A_{0,k}$ is the initial specific activity at $t=0$ (units: Bq/kg), and $T_{1/2,k}$ is the decay half-life.
$\varepsilon_{k,i}$ is the detection efficiency of the decay produts of $k$ isotope in $i$ energy range 
(units: cpd/(Bq/kg), cpd as count per day).
$N_{Iso}$ is the total number of cosmogenic isotopes.
The second item $\sum_{p}C_{p,i}\cdot \varepsilon_{p,i}$ is the sum of the contribution from radon daughters. 
In this work, we treat each radon daughter independently as they are not in equilibrium, but make the assumption
that they are uniformly distributed in the measurement chamber.
$C_{p}(t)$ is the average concentration of radon daughter $p$ (units: Bq/m$^3$) in $t\sim t+\Delta t$ interval.
$\varepsilon_{p,i}$ is the detection efficiency of $\gamma$-rays emitted by $p$ isotope in $i$ energy range (units: cpd/(Bq/m$^3$)).
$N_{Rn}$ is the total number of considered radon daughters.
$B_i$ is the time-invariant component in $i$ energy range (units: cpd).

We built the detector and its shielding structure in Geant4\cite{bib:7,bib:8,bib:9} to simulate
the detection efficiency of different background sources.
The geometry in Geant4 includes the crystal, its holding structure, the cryostat, and the shielding.
The simulation is performed using Geant4 version 11.0.3 with the $shielding$ physics list.
In the simulation, the energy deposition in the surface dead layer is not recorded, 
and the energy resolution of the detector is considered using the calibrated resolution function:
$FWHM /\mathrm{keV}=0.9903+0.3197\sqrt{E/\mathrm{keV}+5.789\times 10^{-5}\times (E/\mathrm{keV})^2}$ \cite{bib:16}.
The efficiencies $\varepsilon_{k,i}$ and $\varepsilon_{p,i}$ are then calculated by:

\begin{eqnarray}
    \label{eq:2}
    \varepsilon_{k,i}=\frac{n_i}{N_S}\cdot m_{Ge},\\
    \label{eq:3}
    \varepsilon_{p,i}=\frac{n_i}{N_S}\cdot V_{Chamber},
\end{eqnarray}

\noindent
where $N_S$ is the total number of simulated particles, 
$n_i$ is the observed count in $i$ energy range in the simulated spectrum,
$m_{Ge}$=2.48 kg is the mass of Ge crystal,
$V_{Chamber}$=9.4 dm$^3$ is the volume of the detector chamber inside the copper shielding.

The activity of cosmogenic isotope ($A_{0,k}$) and radon daughter concentration ($C_{p}(t)$) 
can be calculated by fitting the modeled count rate to measured data using a maximum likelihood method.
The observed count in each time interval obeys the Possion distribution, 
and the likelihood function can be written as:

\begin{eqnarray}
    \label{eq:4}
    \mathcal{L}=\prod_{i}^{N_E}\prod_{t}^{N_T}\frac{\lambda_{i,t}^{n_{i,t}}}{n_{i,t}!}e^{-\lambda_{i,t}},\\
    \label{eq:5}
    \lambda_{i,t}=R_i(t,t+\Delta t)\cdot \Delta t,
\end{eqnarray}

\noindent
where $\lambda_{i,t}$ and $n_{i,t}$ is the expected and observed count in $i$ energy range at $t \sim t+\Delta t$ interval.
$N_E$ is the number of selected energy intervals, and $N_T$ is the number of time intervals.
The total number of fitting parameters is $N_{Iso}+N_{Rn}\cdot N_T+N_E$, 
corresponding to the number of cosmogenic isotopes, radon daughter concentration, and time-invariant background.
And $N_E\cdot N_T$ is the number of observed data, corresponding to the count rate series in each energy range.
To better constrain the fitting parameters, multiple energy ranges are required, they should at least include
the signal of cosmogenic isotopes and characteristic peak of radon daughter.
It should be noted that the fitting of the radon daughter concentration $C_p(t)$ is only sensitive to
the variation of its concentration, the constant part will be regarded as a time-invariant background.

In the calculation of Ge-68 activity,
we select three energy ranges: 609$\pm$5 keV, 1764$\pm$6 keV, and 1000$\sim$3000 keV,
the first two are characteristic peaks of the radon daughter Bi-214, 
and the third is chosen as the signal region of Ge-68 and its decay daughter Ga-68.
We assume the Ge-68 and Ga-68 are in equilibrium, and hereafter use Ge-68 to indicate Ge-68 and Ga-68.
We only consider two time variant-components: Ge-68 and the airborne radon daughter Bi-214.
Based on activities determined in our previous work\cite{bib:15}, 
other cosmogenic isotopes (Mn-54, Co-57, and Co-58) contribute negligible background in the selected energy regions
(less than 1\%), therefore are omitted in this work. 
The total 358 spectra measured in a 6-hour interval are merged into 30 spectra,
each corresponding to 3 days of live measurement time.
The count rate in each energy range is calculated for the 30 spectra, 
and a total of 90 count rate data are used for analysis.

The fitting of the parameters is by maximum the likelihood function $\mathcal{L}$ in Eq.\ref{eq:4}.
We use the Markov Chain Monte Carlo (MCMC) method\cite{bib:19} to calculate 
the best-fit result and the corresponding uncertainty.
The calculation is performed using the $UltraNest$ MCMC toolkit\cite{bib:17} in the $python3$ platform.

\subsection{Statistic test of the significance of Ge-68 signal}{\label{sec.2.2}}
In order to evaluate the significance of the Ge-68 signal, we perform a hypothesis test:
the null hypothesis ($H_0$) is that there are no Ge-68 in the measured data, 
and the alternative hypothesis ($H_1$) is the best-fit result.
For the null hypothesis, Bi-214 concentration and time-invariant background are 
fitted using the same procedure while setting the Ge-68 activity to 0.

The $P$ value is used to test the consistency between the hypothesis and measurement.
It is defined as the probability of getting a worse result than observed in the measurement 
under a specified hypothesis\cite{bib:18}.
In this work, it is calculated by:

\begin{eqnarray}
    \label{eq:6}
    P(H_i)=\int_{-\infty}^{s_{obs}(H_i)} f(s|H_i)\mathrm{d}t \quad (i=0,1),\\
    \label{eq:7}
    s(H_i)=\sum_{t}^{N_T}n_{t}\cdot \lambda_{t}(H_i) - \lambda_{t}(H_i)-\ln(n_{t}!),
\end{eqnarray}

\noindent
where $s(H_i)$ is the test statistic defined as the sum of the logarithmic likelihood value in the 1-3 MeV Ge-68 signal region.
$\lambda_{t}(H_i)$ is the expected count in $t\sim t+\Delta t$ interval under $H_i$ hypothesis.
$s_{obs}(H_i)$ is the observed value of $s(H_i)$, and is calculated via Eq.\ref{eq:7} using measured data.
$f(s|H_i)$ is the probability distribution function (PDF) of test statistic $s$ under the $H_i$ hypothesis.

$f(s|H_i)$ is calculated via a toy Monte Carlo method: 
$n_t$ are randomly sample by Possion distribution using $\lambda_{t}(H_i)$ as the expectation.
$s(H_i)$ is then calculated and stored for a group of sampled $n_t$ ($t=1,2,...,N_T$) as a hypothesis experiment.
Then the hypothesis experiment is performed 20,000 times to get the PDF of $s(H_i)$.

The $P$ value of the alternative hypothesis ($P(H_1)$) demonstrates the goodness of the fit,
while the $P$ value of the null hypothesis ($P(H_0)$) indicates if adding Ge-68 in fitting is necessary.
The significance of the Ge-68 signal is the combination of a small $P(H_0)$ and a large $P(H_1)$.

\subsection{Evaluation of the minimum detection activity}{\label{sec.2.3}}
The minimum detection activity (MDA) of a $\gamma$ spectrometer 
relies on the background level, the measured time, the analyzed characteristic peak, and the detection efficiency.
If other factors are kept constant,
the decay of Ge-68 will lead to a decrease in background and an improvement of the MDA.

Here we consider a typical "paired measurement" to evaluate the improvement of MDA by the decrease of Ge-68 background.
In a paired measurement, the sample and background are measured for the same time ($t_m$),
and the MDA is written as\cite{bib:18-2}:

\begin{eqnarray}
    \label{eq:8}
    MDA=\frac{2.71+4.65\sqrt{b_i(t_U,t_m)}}{\varepsilon \cdot I_{\gamma} \cdot t_m},
\end{eqnarray}

\noindent
where $i$ indicates the selected characteristic peak,
$t_U$ is the underground operation time before the sample measurement,
$I_{\gamma}$ is the yield of the characteristic $\gamma$ line,
$\varepsilon$ is the detection efficiency,
$b_i(t_U,t_m)$ is the background count in the analysis window 
($E_i\pm 3\sigma_{E_i}$), $E_i$ is the energy of thecharacteristic $\gamma$ line
$\sigma_{E_i}$ is the energy resolution.
Background $b_i(t_U,t_m)$ is calculated using the 90-day background spectrum 
and the fitted Ge-68 activity ($A_{0,\mathrm{Ge-68}}$) via:

\begin{eqnarray}
    \label{eq:9}
    b_i(t_U,t_m)=&&R_{i,others}\cdot t_m \\ \nonumber
    &&+A_{0,\mathrm{Ge-68}}\cdot \frac{T_{1/2}}{\ln 2}\cdot e^{-\frac{\ln 2}{T_{1/2}}t_U}
    [1-e^{-\frac{\ln 2}{T_{1/2}}t_m}]  \cdot \varepsilon_{\mathrm{Ge-68},i},
\end{eqnarray}

\noindent
where $R_{i,others}$ is the measured background rate in $i$ characteristic peak region 
subtracting the background from Ge-68 using Eq.\ref{eq:1}.
$\varepsilon_{\mathrm{Ge-68},i}$ and $T_{1/2}$ are the detection efficiency and half-life of Ge-68.

Here we set measure time $t_m$ to 30 days, 
and the improvement of MDA after underground operation time ($t_U$) can be written as 
the ratio of MDA between $t_U$ and $t_U=0$:

\begin{eqnarray}
    \label{eq:10}
    \frac{MDA(t_U)}{MDA(t_U=0)}=\frac{2.71+4.65\sqrt{b(t_U,t_m)}}{2.71+4.65\sqrt{b(t_U=0,t_m)}}
\end{eqnarray}

\section{Result and discussion}{\label{sec.3}}
\subsection{Detection efficiency of Ge-68 and radon daughter Bi-214}
\begin{figure}[!htb]
    \centering
    {\includegraphics[width=1\linewidth]{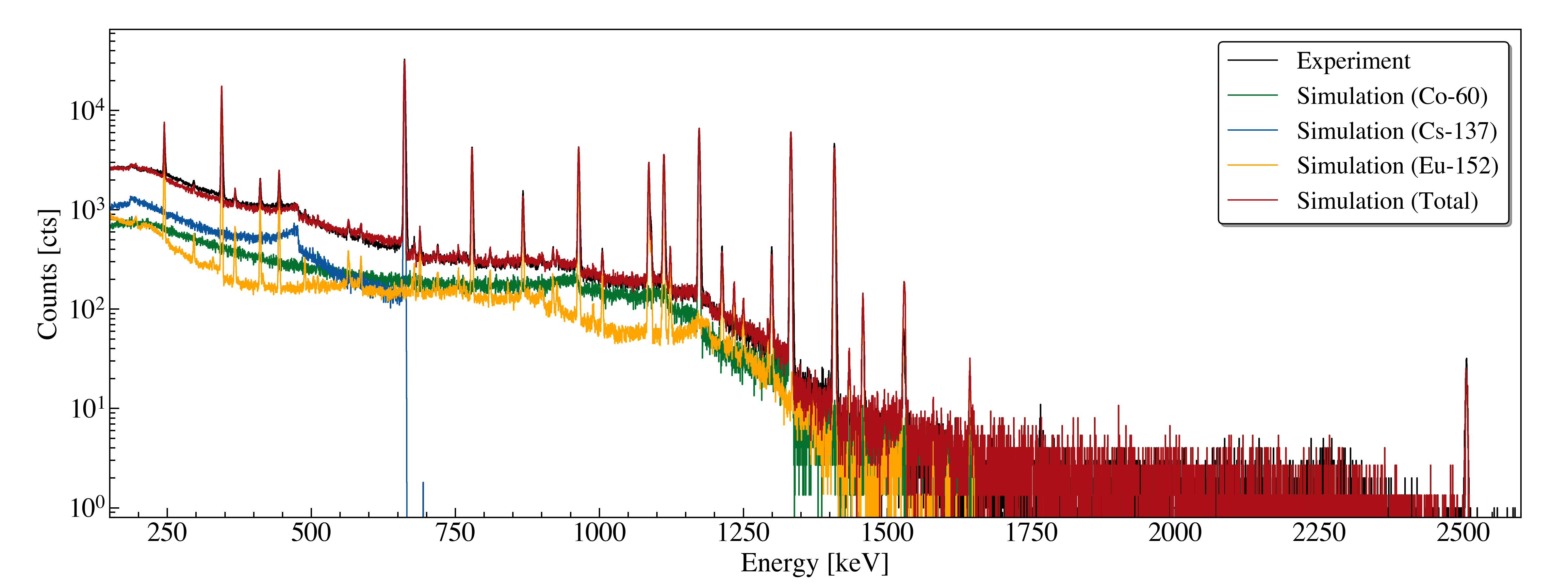}}
    \caption{\label{fig:a1} 
    Comparison between simulated and measured spectra in the calibration experiment.
    The calibration is performed with Co-60, Cs-137, and Eu-152, their simulated spectra are in 
    green, blue, and orange, respectively. The red spectrum is the sum of the simulated spectra.
    The measured spectrum is labeled in black.
    }
\end{figure}

{\label{sec.3.1}}
The detection efficiency is calculated via Geant4 simulation.
To validate our simulation, the spectrum of the calibration experiment is compared with that obtained by the simulation.
The calibration places three radioactive sources, Co-60, Cs-137, and Eu-152 at the top of the detector,
The spectra of the three sources are simulated separately and then added together to compare with the measurement.
As shown in in Fig.\ref{fig:a1}.
the simulated spectrum is in good agreement with the measured one.

\begin{figure}[!htb]
    \centering
    {\includegraphics[width=0.75\linewidth]{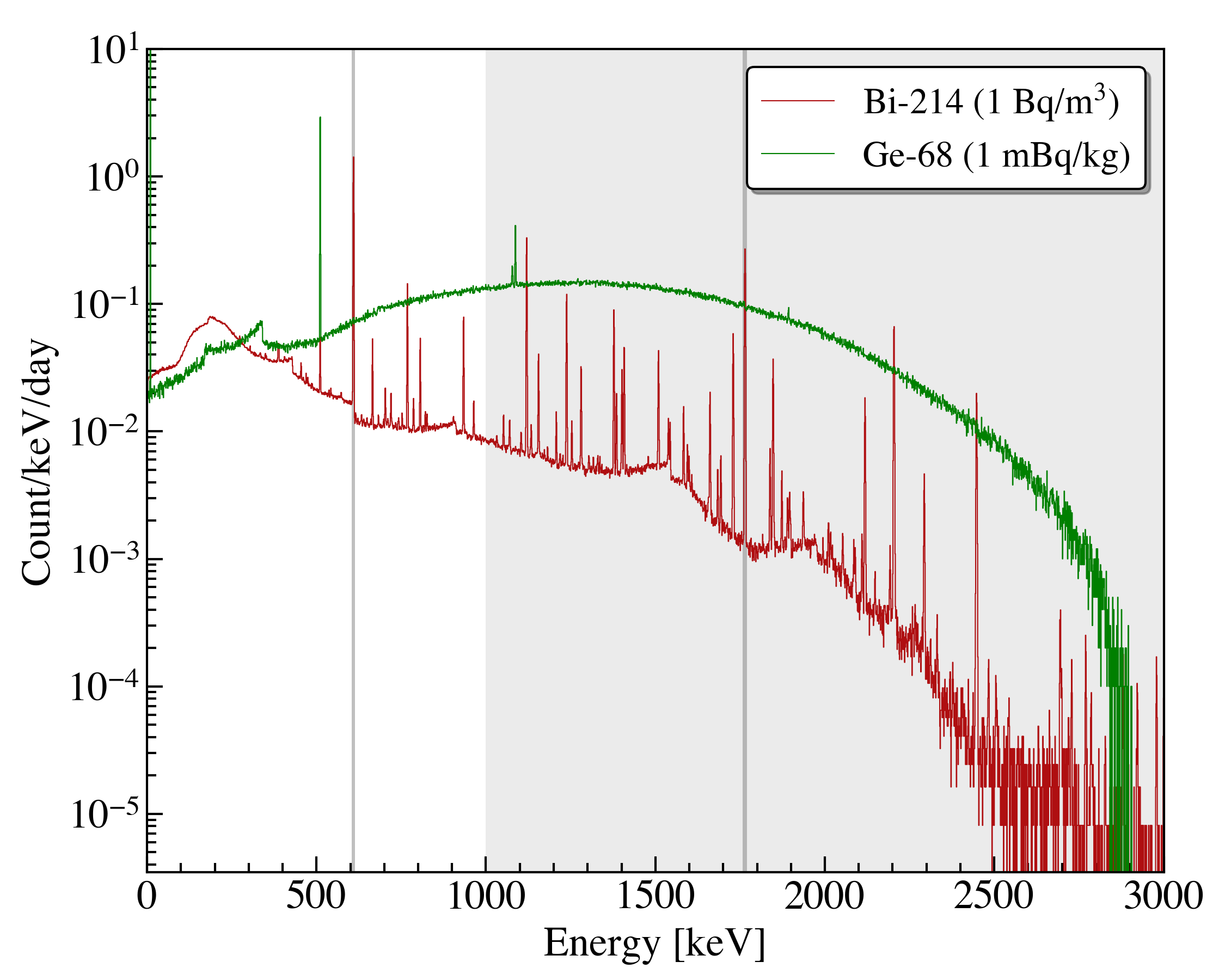}}
    \caption{\label{fig:3} The simulated spectra of 1 mBq/kg Ge-68 and 1 Bq/m$^3$ radon daughter Bi-214.
    The Ge-68 spectrum includes the contribution from its decay daughter Ga-68 under the equilibrium assumption.
    The three selected energy windows are labeled in gray.}
\end{figure}

Total $10^7$ and $10^8$ particles are simulated for calculating the detection efficiencies of Ge-68 and airborne Bi-214,
the simulated spectra are shown in Fig.\ref{fig:3}.
For airborne Bi-214, its detection efficiencies in 609$\pm$5 keV, 1764$\pm$6 keV, and 1000$\sim$3000 keV energy ranges are
3.1, 0.73, and 7.6 cpd/(Bq/m$^3$), respectively.
For Ge-68, the efficiencies are 0.71, 1.20, and 132.3 cpd/(mBq/kg), respectively.
The statistical uncertainties in the simulated detection efficiencies are less than 0.5\% 
and therefore omitted in the analysis. 
The dead layer at the detector surface is set to 0.5 mm according to the manufactory, 
and its uncertainty is not considered in this work.

The count rate from 1 mBq/kg Ge-68 equals that from 17.4 Bq/m$^3$ airborne Bi-214 in 1-3 MeV energy region.
The saturated activity of Ge-68 at sea level is around 1 mBq/kg\cite{bib:10}.
This indicates that despite the spectrum of Ge-68 has no significant characteristic peak, 
it still can contribute a significant background in the total spectrum.

\subsection{Fitting result of Ge-68 activity}
\begin{figure}[!htb]
    \centering
    {\includegraphics[width=1.0\linewidth]{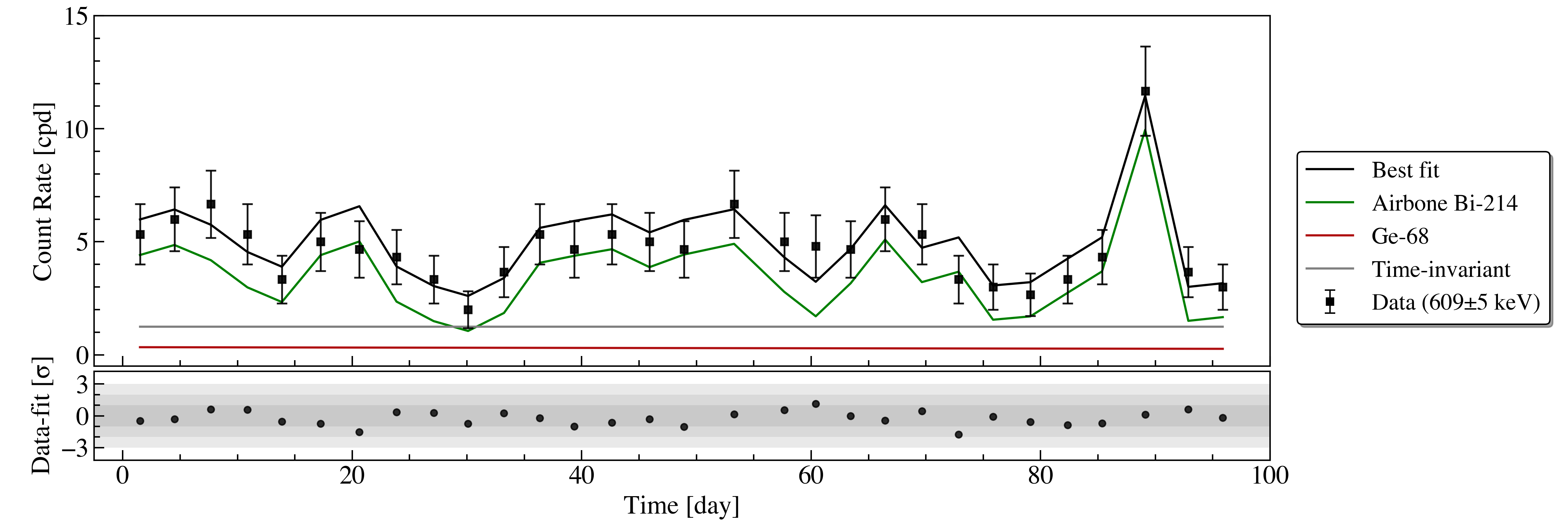}}
    {\includegraphics[width=1.0\linewidth]{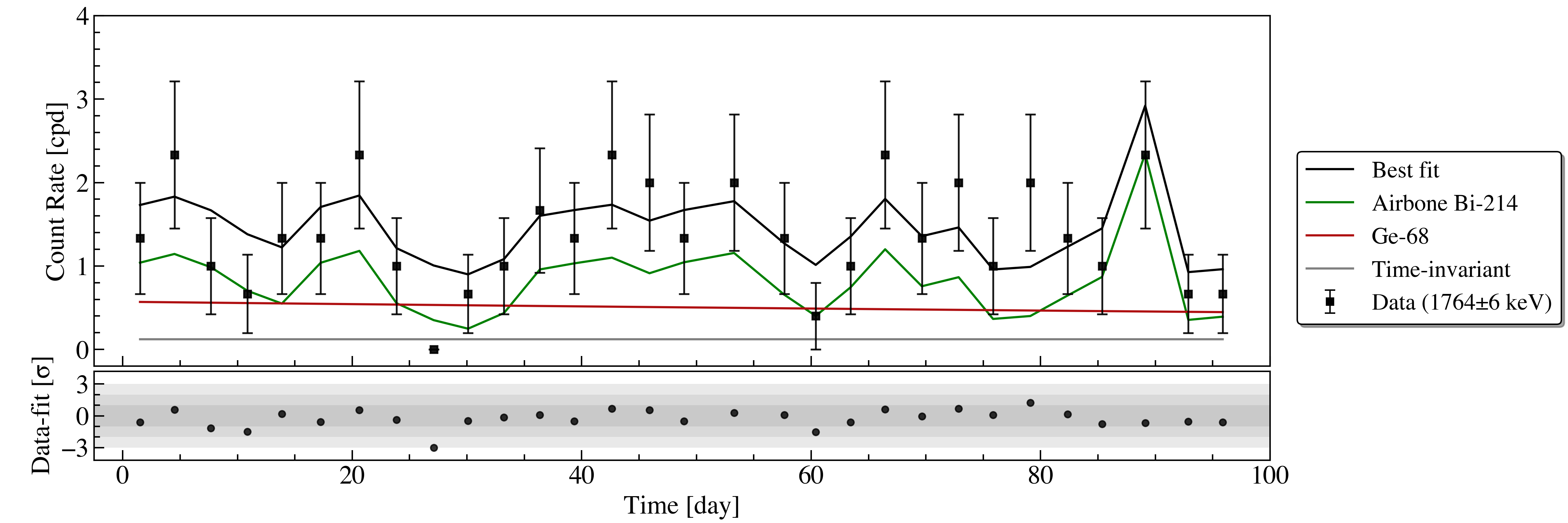}}
    {\includegraphics[width=1.0\linewidth]{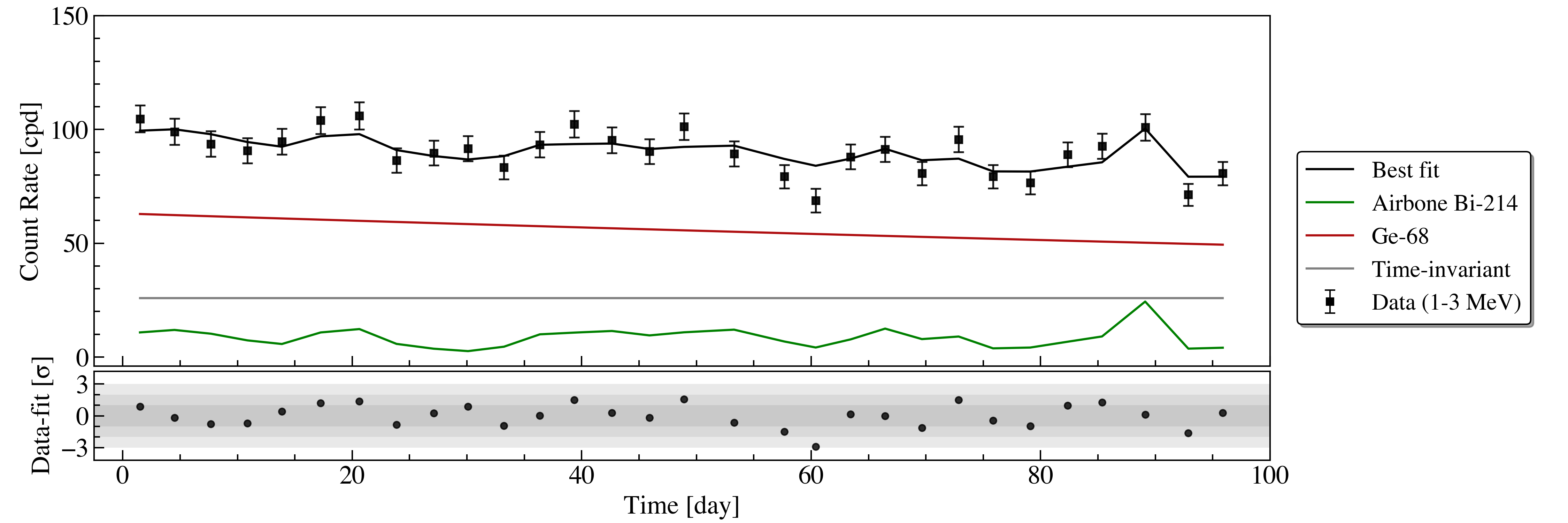}}
    \caption{\label{fig:4} The best-fit result of the count rate in 609$\pm$5 keV, 1764$\pm$6 keV, and 1000$\sim$3000 keV.
    The best-fit curves are labeled in black, and the fitted contributions from Ge-68, airborne Bi-214, 
    and time-invariant background are labeled in red, green, and gray respectively. 
    The residuals between the best-fit and measured data are shown below the best fit curves.}
\end{figure}

{\label{sec.3.2}}
The count rates in 609$\pm$5 keV, 1764$\pm$6 keV, and 1000$\sim$3000 keV energy ranges of the 30 measured spectra
are fitted with contributions from Ge-68, radon daughter Bi-214, and time-invariant background.
The fit results are shown in Fig.\ref{fig:4}
and are in good agreement with the measured data.
The residuals between fitted and measured data are around 0 
and mostly within the 3$\sigma$ band of the statistical uncertainties in measured data.
Our method decouples the contributions from Ge-68 and radon daughter Bi-214 as in Fig.\ref{fig:4},
the 609$\pm$5 keV and 1764$\pm$6 keV region are dominated by the contribution of Bi-214
and provide a strong constraint to the Bi-214 concentration during the measurement.
In the 1-3 MeV region, the contribution from Bi-214 is about 9.4\% of the total count rate,
Ge-68 contributes about 62.0\% of the total count rate and is the most dominant background source.

\begin{figure}[!htb]
    \centering
    {\includegraphics[width=0.47\linewidth]{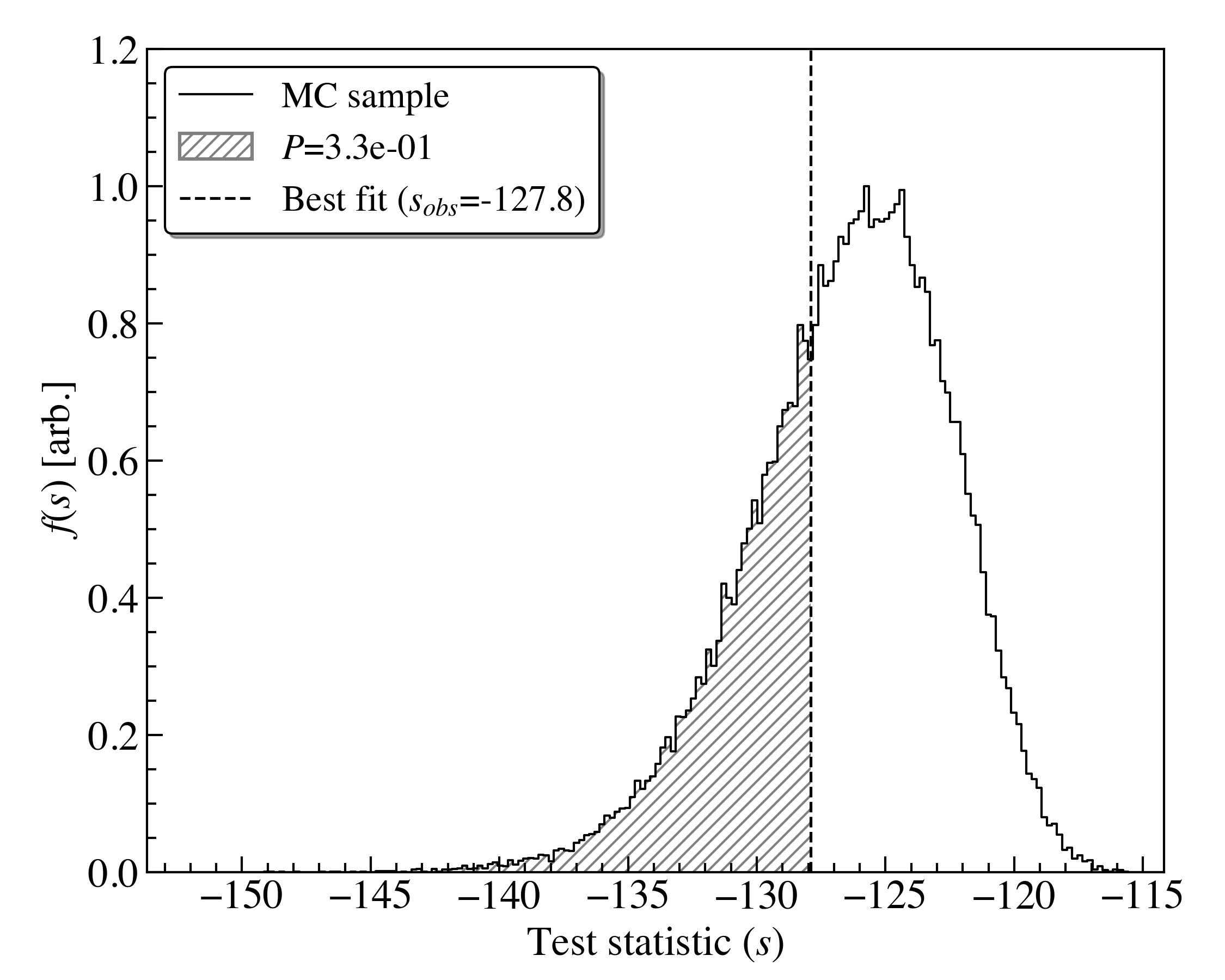}}
    {\includegraphics[width=0.47\linewidth]{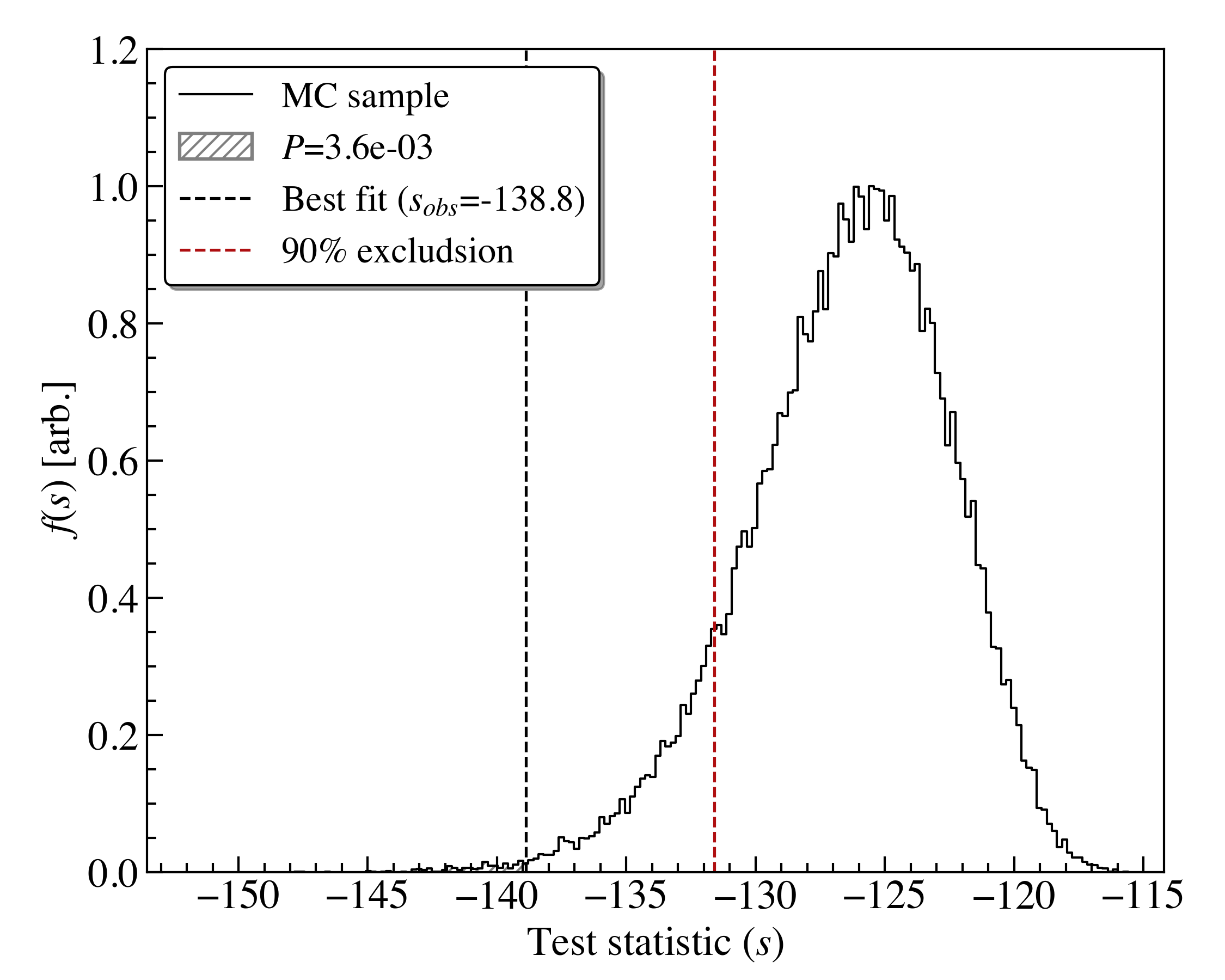}}
    \caption{\label{fig:5} 
    The significance test of Ge-68 signal. The left panel is the
    PDF of test statistic $s$ under alternative hypothesis (the best fit result in Fig.\ref{fig:4}),
    The right panel is the PDF of $s$ under null hypothesis (no Ge-68 signal).}
\end{figure}

The significance of the Ge-68 signal is tested using the method described in Sec.\ref{sec.2.2}.
The calculated PDF of the test statistic $f(s|H_i)$ for null and alternative hypothesis are shown in Fig.\ref{fig:5}
alone with the observed value ($s_{obs}$).
the $P$ value of the null hypothesis $P(H_0)$=0.036\%, 
which excludes the no Ge-68 hypothesis at 99.64\% confidence level.
The $P$ value of the alternative hypothesis $P(H_1)$=33\%,
demonstrates the goodness of the fit and indicates a significant Ge-68 signal in the measured data.

\begin{figure}[!htb]
    \centering
    {\includegraphics[width=0.85\linewidth]{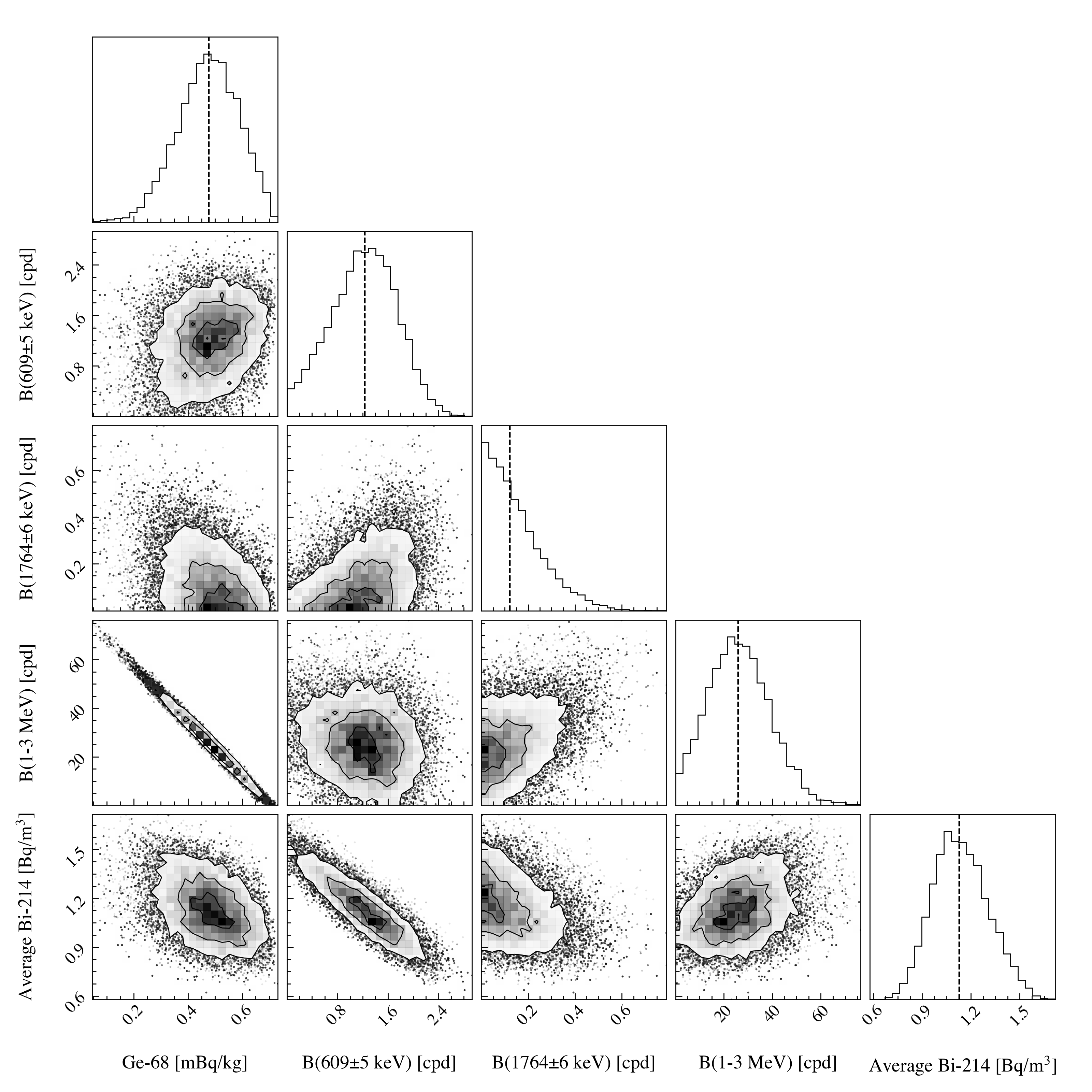}}
    \caption{\label{fig:6} Contours for pairs of parameters in the $UltraNest$ MCMC sampling,
    and the projected likelihood distribution for each parameter.
    The colors indicate the relative value of the likelihood distribution,
    each vertical dash line indicates the median position.
    For simplicity, the Bi-214 concentrations at different times are combined to an average value.}
\end{figure}

The fitted Ge-68 initial activity is 477.0$\pm$112.4 $\mu$Bq/kg, 
the uncertainty corresponding to the 68\% confidence interval 
in the likelihood distribution derived from the $UltraNest$ MCMC sampling.
Values and uncertainties of other parameters are listed in Tabel.\ref{tab:1}.
Fig.\ref{fig:6} demonstrates the contours for pairs of parameters in the MCMC sampling,
which shows the correlation between each pair of parameters.
The largest correlation is between Ge-68 activity and time-invariant background in 1-3 MeV,
which is also the main contributor to the uncertainty of Ge-68 activity.

Theoretical calculations give Ge-68 production rate in a range of 80-200 atom/kg/day at sea level 
for natural germanium detector \cite{bib:10,bib:ad-1}
corresponding to saturated activities of 925-2200 $\mu$Bq/kg. 
The uncertainly in the theoretical calculations mainly arise from the uncertainty of cosmic-ray neutron flux, 
which is affected by factories including altitude, latitude, and shielding. 
Our measured Ge-68 activity ($476.95\pm112.38$ $\mu$Bq/kg) is significantly lower than the saturated activity at sea level. 
During the non-working hours in its fabrication, the detector was temporarily stored in a shallow underground location 
with an overburden exceeding 50 meters of water equivalent. 
The underground location provided shielding against the cosmic-rays and reduced the production rate of Ge-68 in the detector. 
However, due to the unknow exposure history before the detector left the manufactory, 
we cannot calculate the production rate of Ge-68 using the measured Ge-68 activity.
For readers interested in the aclculation and evaluation of Ge-68 production rate,
they can refer to\cite{bib:10,bib:ad-1,bib:ad-II-1,bib:ad-II-2,bib:ad-II-4} for more details.

\begin{table}
    \centering
    \caption{\label{tab:1}
    Fit results of Ge-68 initial activity ($A_{0,\mathrm{Ge-68}}$), time-invariant background ($B$), 
    and Bi-214 concentration ($C_{\mathrm{Bi-214}}$).
    For simplicity, the Bi-214 concentrations at different times are combined to an average value.}
    \renewcommand\arraystretch{1.5}
    \begin{tabular}{cccc}
    \hline
        Parameter & Units & Value & Note\\
    \hline    
        $A_{0,\mathrm{Ge-68}}$ & $\mu$Bq/kg & 476.95$\pm$112.38 & Ge-68 initial activity \\
        $B_{\mathrm{609\pm 5 \thinspace keV}}$ & cpd & 1.24$\pm$0.53 & time-invariant background \\
        $B_{\mathrm{1764\pm 6 \thinspace keV}}$ & cpd & 0.12$\pm$0.12 & time-invariant background \\
        $B_{\mathrm{1\sim 3 \thinspace MeV}}$ & cpd & 25.81$\pm$12.71 & time-invariant background \\
        $C_{\mathrm{Bi-214}}$ & Bq/m$^3$ & 1.42$\pm$0.17 & average Bi-214 concentration \\
        \hline
    \end{tabular}
\end{table}

\subsection{Background induced by Ge-68 and its effects on MDA}
\begin{figure}[!htb]
    \centering
    {\includegraphics[width=1.0\linewidth]{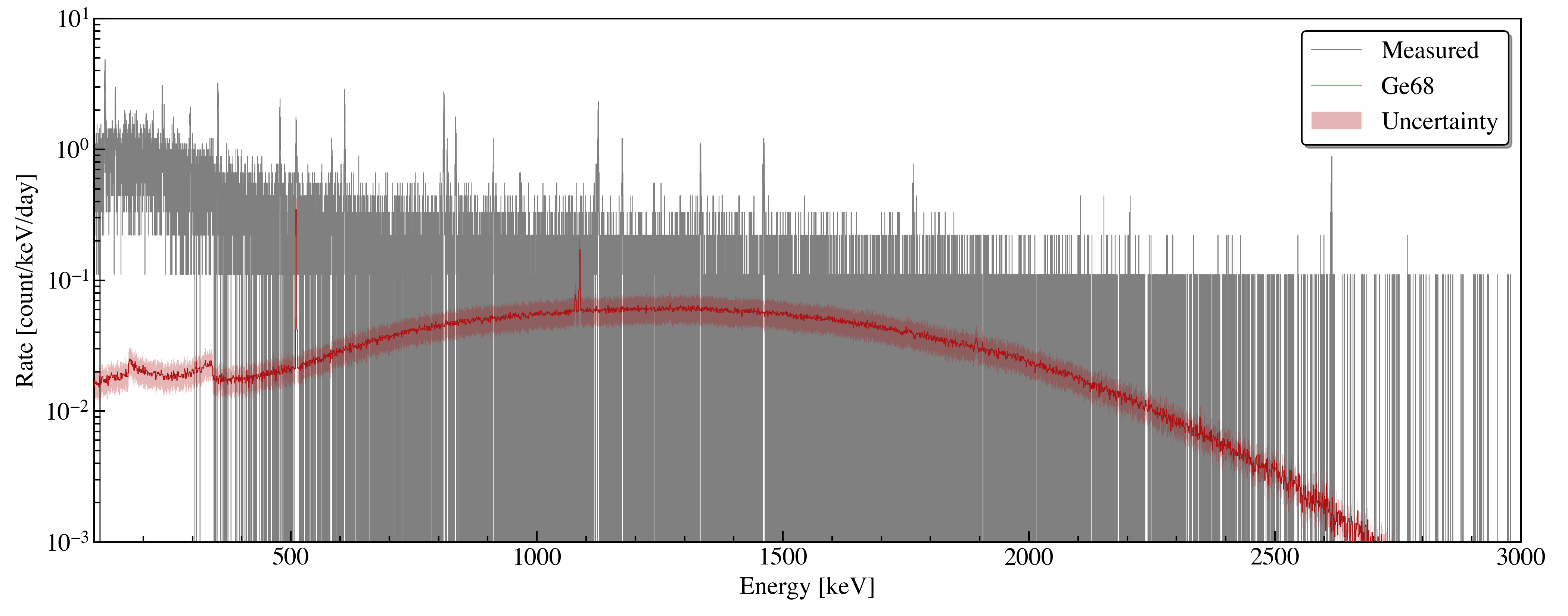}}
    \caption{\label{fig:7} Comparison of measured spectrum (gray line) 
    and simulated Ge-68 (Ga-68) spectrum (red line) with fitted activity.
    The fitted initial activity of Ge-68 is 477.0$\pm$112.4 $\mu$Bq/kg,
    and the red shadow is the spectra corresponding to the fit uncertainty.}
\end{figure}

{\label{sec.3.3}}
Fig.\ref{fig:7} shows the background contribution of Ge-68 and its decay daughter Ga-68 
in the 90 days background spectrum.
As Ge-68 and its decay daughter Ga-68 are in radioactive equilibrium,
we use Ge-68 to indicate Ge-68 and Ga-68 hereafter.
The red line is the spectrum corresponding to 477 $\mu$Bq/kg initial activity,
and the red shadow indicates the $\pm$112.4 $\mu$Bq/kg fit uncertainty.
In 1-3 MeV energy region,
the measured background is 90.9$\pm$1.0 cpd, 
and the Ge-68 contributes 55.9$\pm$13.2 cpd, about 62\% of the measured background.

\begin{figure}[!htb]
    \centering
    {\includegraphics[width=0.48\linewidth]{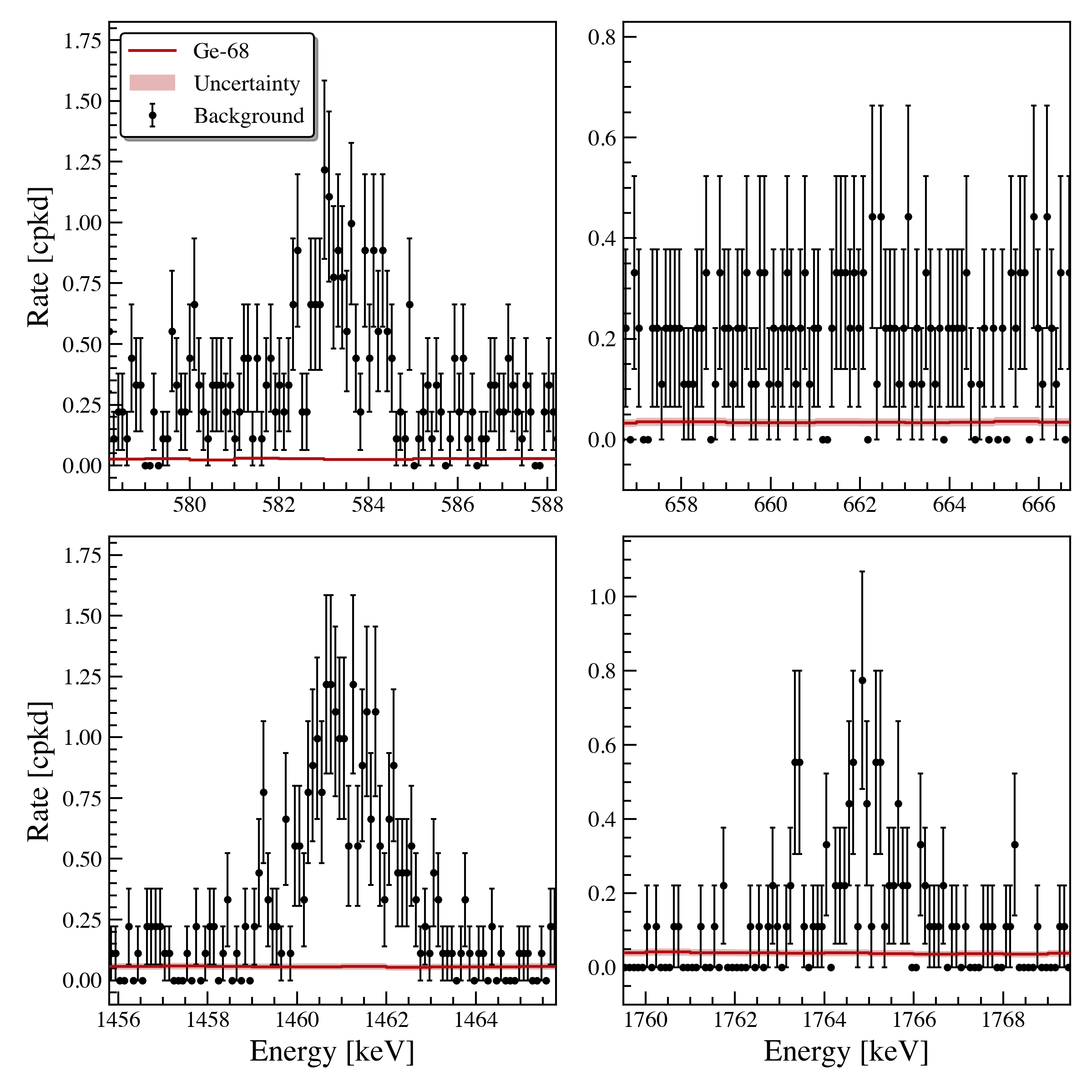}}
    {\includegraphics[width=0.48\linewidth]{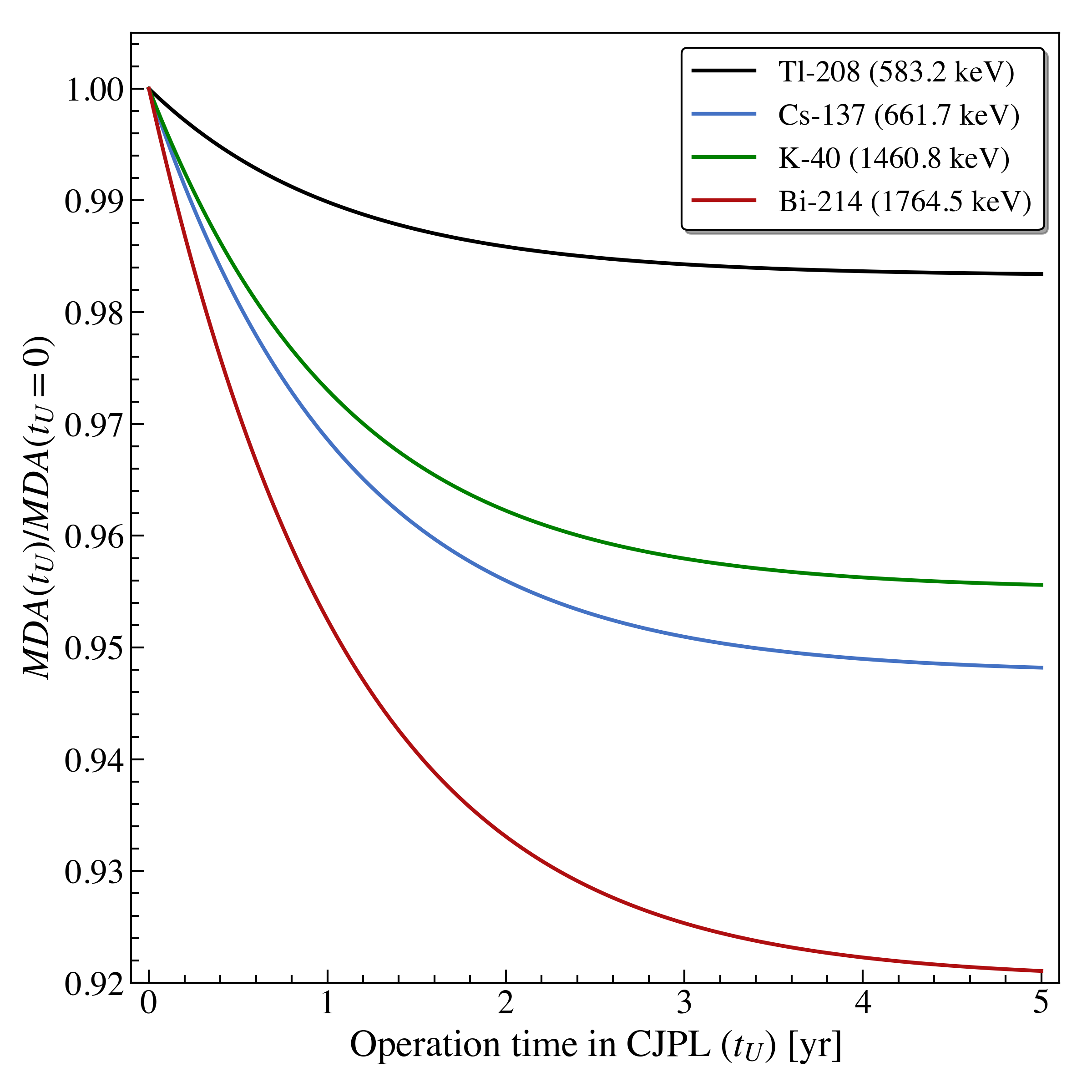}}
    \caption{\label{fig:8} Left: the 90-day background spectrum in the four characteristic peak regions,
    contributions from Ge-68 (Ga-68) are labeled in red, 
    the measurement of the background spectrum started about 21 days after the detector arrived at CJPL.
    Right: the ratio of MDA between $t_U$ and $t_U=0$ for the four characteristic peaks.}
\end{figure}

We select four characteristic $\gamma$ lines from different isotopes 
to evaluate the effects of Ge-68 background on their minimum detection activity (MDA).
The select peaks are:
583.2 keV (Tl-208), 661.7 keV (Cs-137), 1460.8 keV (K-40), and 1764.5 keV (Bi-214).
The 90-day background spectrum of the four characteristic $\gamma$ peaks
and the contribution from Ge-68 (Ga-68) are shown in the left panel of Fig.\ref{fig:8}.
For simplicity, our analysis assumes a constant background rate from the airborne radon daughter 
to better demonstrate the effects from the change of Ge-68 background.
The improvement of MDA at different operation time in CJPL is calculated using Eq.\ref{eq:10},
% the sample measure time $t_m$ is set to 
the results are shown in the right panel of Fig.\ref{fig:8}.

After 5-year operation at CJPL,
the Ge-68 activity decreases from 477.0 $\pm$112.4 $\mu$Bq/kg to 4.47$\pm$1.05 $\mu$Bq/kg
and results in a 2\%$\sim$8\% MDA improvement for the selected four characteristic peaks.
The smallest 2\% improvement is for the Tl-208 583.2 keV peak 
as the background is dominated by Tl-208 in detector structure materials, 
Ge-68 only contributes 4.6\% of the total background at $t_U$=0.
For the Bi-214 1764.5 keV characteristic peak, 
Ge-68 contributes 21\% of the total background at $t_U$=0 and 0.2\% at $t_U$=5 years,
the about 20\% background reduction gives an 8\% improvement in the MDA.

\subsection{Variation of airborne Bi-214 concentration}
\begin{figure}[!htb]
    \centering
    {\includegraphics[width=1.0\linewidth]{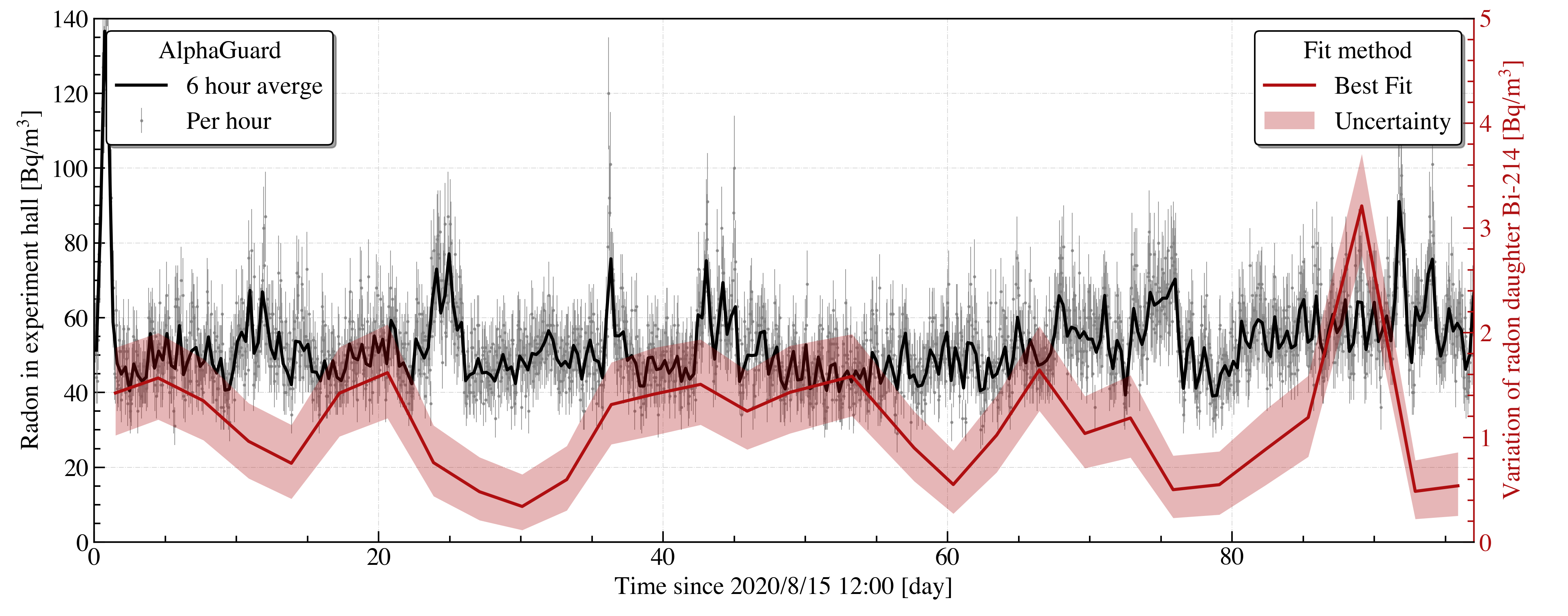}}
    \caption{\label{fig:9} Variation of the radon concentration in the experiment hall and 
    the airborne Bi-214 in the detector chamber within the copper shielding.
    The gray point and black line are the radon concentration measured by an AlphaGuard radon monitor.
    The red line is the fitted airborne Bi-214 concentration and the red shadow is the corresponding fit uncertainty.}
\end{figure}

{\label{sec.3.4}}
Our method also provides the concentration variation of radon daughter Bi-214 in the detector chamber.
As the detector chamber within the copper shielding is constantly purged by nitrogen gas,
a comparison between the variation of Bi-214 in detector chamber and radon in experiment hall
indicates whether air has been mixed in the nitrogen gas or there is a leakage point in the shielding.
The radon (Rn-222) concentration in the experiment hall has been measured by an AlphaGuard PQ2000 radon monitor.
The AlphaGuard is set to the diffusion mode and a measurement period of one hour.
Fig.\ref{fig:9} demonstrates the comparison of the Bi-214 and the Rn-222 concentration.

The variation of airborne Bi-214 is within 5 Bq/m$^3$, 
and its average value (1.42$\pm$0.17 Bq/m$^3$) is about 40 times lower than the average 
Rn-222 concentration in the experiment hall.
There is also no significant coincidence between the Bi-214 and Rn-222 concentrations,
indicating that the detector chamber has been well isolated from the experiment hall.

\subsection{Effects of time-invariant background on fitting sensitivity}
\begin{figure}[!htb]
    \centering
    {\includegraphics[width=1.0\linewidth]{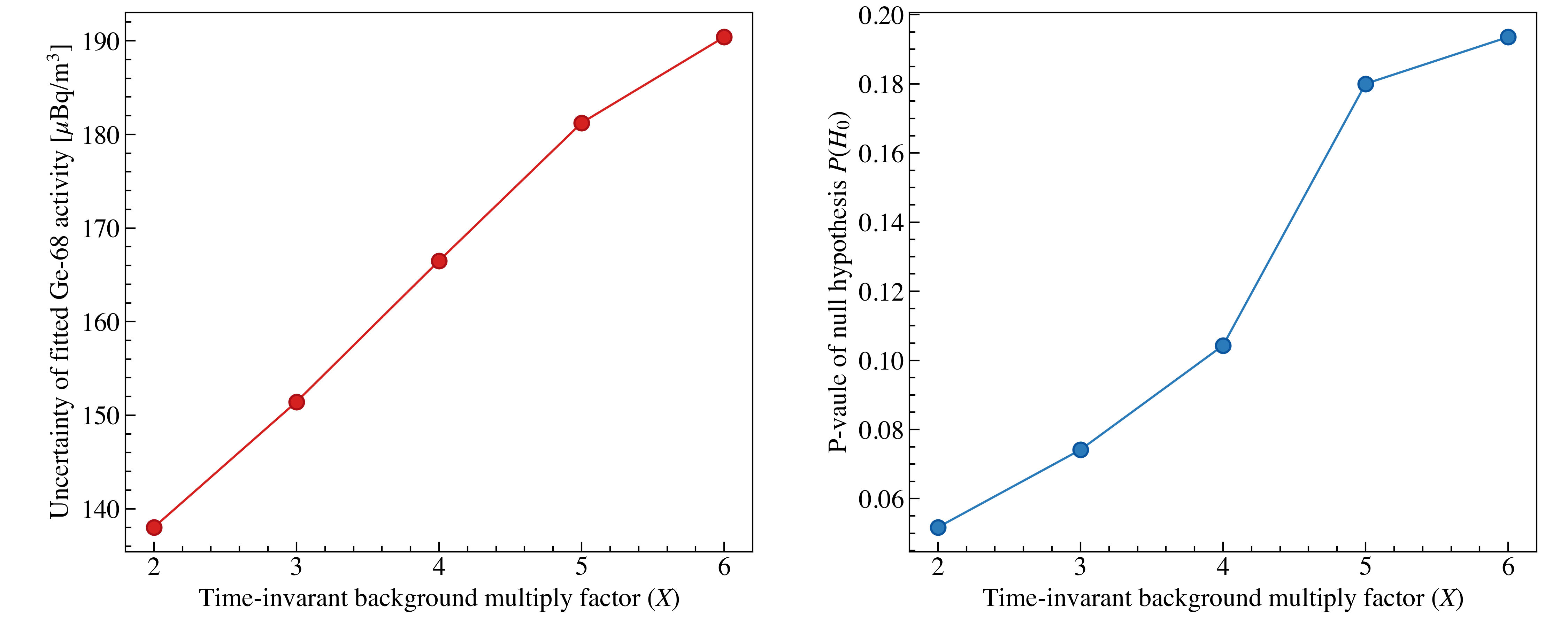}}
    \caption{\label{fig:10} 
    The uncertainty of fitted Ge-68 activity (Left) and the P-value of the null hypothesis (Right) as a 
    function of the factor ($X$). The time-invariant background is multiplied by $X$.
    }
\end{figure}

{\label{sec.3.5}}
The statistical fluctuations of the time-invariant backgrounds could jeopardize the time series fitting.
Therefore, for other HPGe spectrometers with higher background levels,
the sensitivity of our method could be degarded by the time-invariant backgrounds.
To evaluate this effect, we tune our best fit model to generate count rate data with different time-invariant backgrounds.
The count rate  data are generated using Eq.\ref{eq:1} and the best fit values in Table.\ref{tab:1}, 
except that the time-invariant backgrounds are multiply by a factor ($X$). 
Then the generated data are fitted by the time-series fitting method and 
the P-value of the null hypothesis ($P(H_0)$) is calculated using Eq.\ref{eq:6}.
Fig.\ref{fig:10} shows the fit uncertainty of Ge-68 activities and $P(H_0)$ as a function of the factor ($X$). 
As the time-invariant background increases, 
although the fitting parameters are still unbiased estimators,
the uncertainty and P(H0) both increases, and when the factor $X$=4, 
corresponding to a $B_{1-3MeV}$ of 103.2 cpd, the significance of Ge-68 signal is less than 90\%. 
It demonstrates that the control of the time-invariant background is a key to improve the sensitivity of our method.

\section{Conclusions}{\label{sec.4}}
In this work, we develop a time series fitting method to calculate the activity of cosmogenic Ge-68 
in a coaxial HPGe detector operated at China Jinping underground laboratory.
Our method uses the change of count rate in 1-3 MeV energy region 
and characteristic peaks of airborne radon daughter to decouple the Ge-68, radon daughter, and time-invariant background
and fit the Ge-68 activity.
The simulated detection efficiencies in different energy regions are used to connect the count rate
to Ge-68 activity and airborne Bi-214 concentration.
A total 90-day measurement data are used in the analysis
and the Ge-68 initial activity is fitted to be 477.0$\pm$112.4 $\mu$Bq/kg,
it contributes about 62\% background in the 1-3 MeV energy region.
Based on the measured background spectrum and fitted Ge-68 activity,
we predict the minimum detection activity for four radioisotopes (Tl-208, Cs-137, K-40, and Bi-214)
will improve by 2\%$\sim$8\% after 5-year underground operation.
For other low background experiments, for instance, 
experiment using HPGe to search for the Ge-76 neutrinoless double beta decay, 
Ge-68 is also an important background in the 2 MeV signal region\cite{bib:ad-1}.
Therefore, the sensitivity of such experiments could be benefit from the reduction of Ge-68 background.
There are other measures to reduce the Ge-68 background in HPGe detectors,
for instance, optimizing above-ground exposure time, additional shielding during transportation, 
and underground detector fabrication\cite{bib:ad-II-3,bib:ad-II-1,bib:ad-II-2}.

Our method can be extended to other cosmogenic isotopes in germanium, 
for instance, Mn-54, Co-57, and Co-58.
As our method relies on observing a significant decay signal of the cosmogenic isotopes
during the measurement time, 
it will be difficult to measure isotopes with a long half-life (for instance, H-3 and Co-60) using the time-series fitting method.
The fitting result of Ge-68 activity could be 
used as an input to a spectrum fitting method to decouple background from different structure materials.

\section*{Acknowledgments:}
This work was supported by the 
National Key Research and Development Program of China 
(Grant No. 2023YFA1607101, 2022YFA1604701)
and the National Natural Science Foundation of China 
(Grant No. 12425507 and 12175112).
We would like to thank CJPL and its staff for supporting this work. 
CJPL is jointly operated by Tsinghua University and Yalong River Hydropower Development Company.

% \section{Bibliography styles}

% There are various bibliography styles available. You can select the style of your choice in the preamble of this document. These styles are Elsevier styles based on standard styles like Harvard and Vancouver. Please use Bib\TeX\ to generate your bibliography and include DOIs whenever available.

% Here are two sample references: \cite{Feynman1963118,Dirac1953888}.

% \section*{References}

% \bibliography{mybibfile}

\end{document}